# Observation of tunable bandgap and anisotropic Dirac semimetal state in black phosphorus

Jimin Kim[1], Seung Su Baik[2,3], Sae Hee Ryu[1,4], Yeongsup Sohn[1,4], Soohyung Park[2], Byeong-Gyu Park[5], Jonathan Denlinger[6], Yeonjin Yi[2], Hyoung Joon Choi[2,3], Keun Su Kim[1,4]*

[1]Department of Physics, Pohang University of Science and Technology, Pohang 790-784, Korea.

[2]Department of Physics, Yonsei University, Seoul 120-749, Korea.

[3]Center for Computational Studies of Advanced Electronic Material Properties, Yonsei University, Seoul 120-749, Korea.

[4]Center for Artificial Low Dimensional Electronic Systems, Institute for Basic Science, Pohang 790-784, Korea.

[5]Pohang Accelerator Laboratory, Pohang University of Science and Technology, Pohang 790-784, Korea.

[6]Advanced Light Source, E. O. Lawrence Berkeley National Laboratory, CA 94720, USA.

*Correspondence to: keunsukim@postech.edu

Black phosphorus consists of stacked layers of phosphorene, a two-dimensional semiconductor with promising device characteristics. We report the realization of a widely tunable bandgap in few-layer black phosphorus doped with potassium using an *in-situ* surface doping technique. Through band-structure measurements and calculations, we demonstrate that a vertical electric field from dopants modulates the bandgap owing to the giant Stark effect and tunes the material from a moderate-gap semiconductor to a band-inverted semimetal. At the critical field of this band inversion, the material becomes a Dirac semimetal with anisotropic dispersion, linear in armchair and quadratic in zigzag directions. The tunable band structure of black phosphorus may allow great flexibility in design and optimization of electronic and optoelectronic devices.

Two-dimensional (2D) atomic crystals have continued to show great potential for application in nanoscale devices (*1*). A key issue is controlling their electronic states to overcome the limit of natural properties. Graphene's effectively massless state of charge carriers is a result of the conical band structure (*2*). However, the lack of a bandgap ($E_g$) limits the on-off current ratio in planar field-effect transistors (*3, 4*), and it has been difficult to reliably achieve a sizable $E_g$ without degrading its electronic quality (*5–7*). It would thus be desirable to realize a 2D system with a widely-tunable $E_g$.

A potential candidate is few-layer black phosphorus (BP), a layered material of elemental phosphorus (*5–22*). The single-layer BP (or phosphorene) has a honeycomb network similar to graphene, but is strongly puckered (armchair-shaped along *x* and zigzag-shaped along *y* in Fig. 1A), rendering its electronic state highly susceptible to external perturbations (*11–22*). The low-energy band structure of phosphorene can be approximated by a bonding and anti-bonding pair of mainly $3p_z$ orbitals (*11, 12*). The corresponding valence band (VB) and conduction band (CB) are located at the zone center ($\Gamma_2^+$ and $\Gamma_4^-$ states in Fig. 1B) with the predicted $E_g$ of 0.7 ~ 1.6 eV (*13–17*). For multilayers, the introduction of interlayer coupling reduces $E_g$ with increasing film thickness, and reaches about 0.33 eV in bulk BP (*11–14*). The $E_g$ of BP films and nanoribbons has been widely predicted to be tunable by strain (*15–17*) and electric field (*17–21*), the latter of which is more viable in gated devices. The electric field affects the real-space distribution of VB and CB states to be shifted in opposite directions to each other. Their potential difference and band mixing lead to a reduction in $E_g$, which is often termed the giant Stark effect (*23–26*). However, despite its potential importance for device applications, little is known experimentally about this effect on the surface of 2D semiconductors under a vertical electric field.

Here, we report the realization of a widely tunable $E_g$ in BP by means of the in-situ deposition of potassium (K) atoms, the well-known technique to induce doping and electric field in 2D van-der-Waals systems (*27*). The K atoms on BP donate charges to few top phosphorene layers, which are confined to form 2D electron gas near the surface (Fig. 1A, bottom). The band structure measured by angle-resolved photoemission spectroscopy (ARPES) at low K density is slightly *n*-doped with $E_g$ greater than 0.6 eV (Fig. 1C). With increasing dopant density, the electric field from the ionized K donors gradually reduces $E_g$ owing to the giant Stark effect, as supported by our density-functional-theory (DFT) calculations. Consequently, the electronic state of BP is widely and continuously tuned from a moderate-gap semiconductor to a band-inverted semimetal. At the critical dopant density of this band-inverted transition (*21, 22*), it turns into an anisotropic Dirac semimetal state (Fig. 1D). This control mechanism of $E_g$ should work in dual-gate BP devices for precisely balancing between high mobility and moderate $E_g$.

Figure 1, E to H, shows a series of ARPES spectra taken along the armchair direction $k_x$ with different dopant density near direct $E_g$. As expected for pristine BP (*28, 29*), in Fig. 1E there is a well-defined VB with a nearly parabolic dispersion and with a vertex at 0.15 eV below the Fermi energy $E_F$ (*30*). Assuming the reported $E_g$ of 0.33 eV (*12*), our pristine sample is slightly hole-doped, which explains *p*-type conduction in transport (*12*). We use a *k-p* perturbation formula (*31*), widely accepted to fit the band dispersion of narrow-gap semiconductors, to quantify the hole effective mass. The best fit (white line overlaid) yields $m_x^* = 0.08 \pm 0.03 m_e$, where $m_e$ is the electron rest mass, in good agreement with theoretical calculations (*12, 14*).

Upon electron doping by the K deposition on the surface, the overall band structure rigidly shifts down towards high energies (fig. S2), so that the magnitude of $E_g$ can be directly measured by ARPES. The energy shift of surface bands is accompanied by steep band bending towards the bulk (along *z* in Fig. 1A) to form 2D electronic states in few phosphorene layers, as confirmed by little $k_z$ dependence (*31*). At low dopant density just before the CB minimum drops below $E_F$ (Fig. 1F), VB rigidly shifts down, and $E_g$ can be estimated at about 0.6 eV or slightly greater. This $E_g$ of 2D electronic states is smaller than that predicted for monolayer phosphorene (*13–17*), and comparable to those predicted for few-layer phosphorene, 0.3 ~ 1.0 eV (*13, 14*). With further increasing dopant density, this $E_g$ gradually reduces (Fig. 1G), and becomes zero (Fig. 1H), indicative of a semiconductor-semimetal transition. At the critical dopant density for this transition, where VB and CB touch each other, the band dispersion in the armchair direction $k_x$ becomes linear (Fig. 1H), whereas that in the zigzag direction $k_y$ remains nearly parabolic (fig. S3).

The linear dispersion in $k_x$ can be more clearly identified in high-resolution ARPES data (Fig. 2A). The peak positions follow X-shaped linear bands of VB (red) and CB (blue) with a crossing point (Fig. 2C) that is also revealed in the intensity map at the crossing energy (Fig. 2E). The spectral simulation with linear bands and finite broadening (*31*) (Fig. 2B) successfully reproduces the experimental data in Fig. 2A.

This suggests that BP at the critical density has a spin-degenerate Dirac state as in graphene (*3*) and Dirac semimetals (*21, 32−34*), which can be stabilized by crystalline symmetry, as explained below. A linear fit yields the velocity of charge carriers $v_x = 5.1 \pm 0.9 \times 10^5$ m/s, which is about half of that in graphene (*2*). Shown in Fig. 2, D to G, is a series of constant-energy intensity maps. The ARPES intensity of band contours is modulated by the matrix-element effect that reflects the symmetry of constituent wavefunctions in real-space (*5*). The maps at above and below the energy of the Dirac point $E_D$ (Fig. 2, D and F) show strong suppression along $k_y$ with respect to $k_x$. The resultant two-fold symmetry of intensity patterns confirms their BP origin rather than K. Taking this matrix-element effect into account, the band contours are oval-shaped with the aspect ratio $r \sim 1.9$, which is related to armchair-zigzag anisotropy in Fig. 1A. Around the contour, the band dispersion gradually changes from linear along $k_x$ to quadratic along $k_y$ (fig. S4). These results support the picture of the anisotropic band crossing (Fig. 1D), which is reproduced by DFT calculations as shown below.

In order to systematically follow the evolution of the band structure, we display 3D representation of ARPES spectra as a function of dopant density $N$ in Fig. 3A. The constant-energy cut at $E_F$ shows the Fermi momentum $k_F$ of CB steadily increases. We estimate the electron concentration $n$ based on Luttinger's theorem as $\pi k_F^2 r$ that corresponds to the area enclosed by oval-shaped contours at $E_F$. We found that $n$ is linearly proportional to $N$ (fig. S6), indicative of monotonic charge transfer from dopants to phosphorene layers. In the constant-momentum cut at $k_x = 0$, red diamonds and blue circles are positions of the VB maximum ($E_v$) and CB minimum ($E_c$), respectively. At the initial stage of doping, the $\Gamma_2^+$ state at $E_v$ has rapidly shifted down until the $\Gamma_4^-$ state at $E_c$ drops below $E_F$. As $N$ is increased further, the center energy $E_0 = (E_v + E_c)/2$ slowly shifts down, whereas the $\Gamma_2^+$ and $\Gamma_4^-$ states get progressively closer to each other, and eventually cross at the critical density $N_c = 0.36$ ML ($n = 8.3 \times 10^{13}$ cm$^{-2}$), where band inversion occurs. The magnitude of $E_g$ is calculated as $E_c - E_v$, and plotted as a function of $N$ in Fig. 3B. As can be seen from the figure, $E_g$ is widely and continuously tunable in the range of +0.6 ~ −0.2 eV.

We performed DFT calculations based on four-layer BP with a single K atom on each 2×2 surface unit cell (corresponding to $N \sim 0.4$ ML in experiments). To effectively describe a lower K density, we increase the vertical distance $d$ between K and BP (2.76 Å at equilibrium), such that the effect of K donors is gradually reduced without change in the supercell size (*31*). A series of band calculations as a function of *d* reproduces key aspects of our experimental observations, the variation of $E_g$ and resultant semimetal-semiconductor transition (red circles in Fig. 3B). Furthermore, at around zero $E_g$ (*d* = 3.67 Å), the bands along $k_x$ are linearly dispersing near $E_D$, while those along $k_y$ remain parabolic (Fig. 3C), as observed experimentally. $N_c$ in Fig. 3B corresponds to the critical point of the band-inverted transition, where the topological invariant quantity $Z_2$ switches between 0 and 1 (*21*), induced by electric field instead of spin-orbit interaction. At this quantum critical point, the system is predicted to be in a unique Dirac semimetal state (*32*), in which the band crossing along the rotational symmetry axis (the *y* axis, zigzag) is quadratic, while that along the other axis (the *x* axis, armchair) is linear. Our results thus collectively identify the formation of the anisotropic Dirac semimetal state at $N_c$, resulting from the characteristic puckered structure of phosphorene.

We now discuss the control mechanism of $E_g$. Figure 3D shows partial charge densities of $E_v$ and $E_c$ points in Fig. 3C, separated by a tiny gap less than 10 meV to avoid their degeneracy. The spatial distributions of $E_v$ and $E_c$ states, which are uniform in pristine BP (fig. S9), becomes strongly separated in opposite directions, indicative of electric field generated by the ionized K donors. The $E_c$ states, which have a positive effective mass along *z*, freely move towards the positive electrical potential, while the $E_v$ states, which have a negative effective mass along *z*, are pushed within the body of BP layers (*12*). As stated above, this remarkable spatial separation of $E_v$ and $E_c$ states explains the variation of $E_g$ by the giant Stark effect (*17−21*). Indeed, band calculations for four-layer BP under external electric field confirm the similar reduction in $E_g$ (with no change in $E_F$), and the band crossover at the critical field of ~0.19 V/Å. From this value, we quantify the Stark coefficient $S_L$ for four-layer BP as ~3 Å (*24*), which is comparable to those predicted for few-layer BP (*19, 21*) and transition-metal dichalcogenides (*25, 26*). Since $S_L$ is known to increase with film thickness (*19, 21*), the critical field for bulk BP (or thicker BP films) would be smaller than the practical dielectric strengths of $SiO_2$ and *h*-BN. Our work thus demonstrates the giant Stark effect in BP as an efficient control mechanism of $E_g$, which is generally attainable in 2D semiconductors and devices made out of these materials.

**Acknowledgments**

This work was supported by IBS-R014-D1. S.S.B. and H.J.C. acknowledge support from the NRF of Korea (Grant No. 2011-0018306). Computational resources have been provided by KISTI Supercomputing Center (Project No. KSC-2013-C3-062). S.P. and Y.Y. acknowledge support from the National Research Foundation of Korea (grant 2013R1A1A1004778) and Yonsei University Future-leading Research Initiative of 2014 (2014-22-0123). PLS-II was supported in part by MSIP and POSTECH. ALS was supported by the US Department of Energy, Office of Sciences under Contract No. DE-AC02-05CH11231. We thank S. W. Jung, W. J. Shin, Y. K. Kim, B. Kim for help in ARPES experiments.

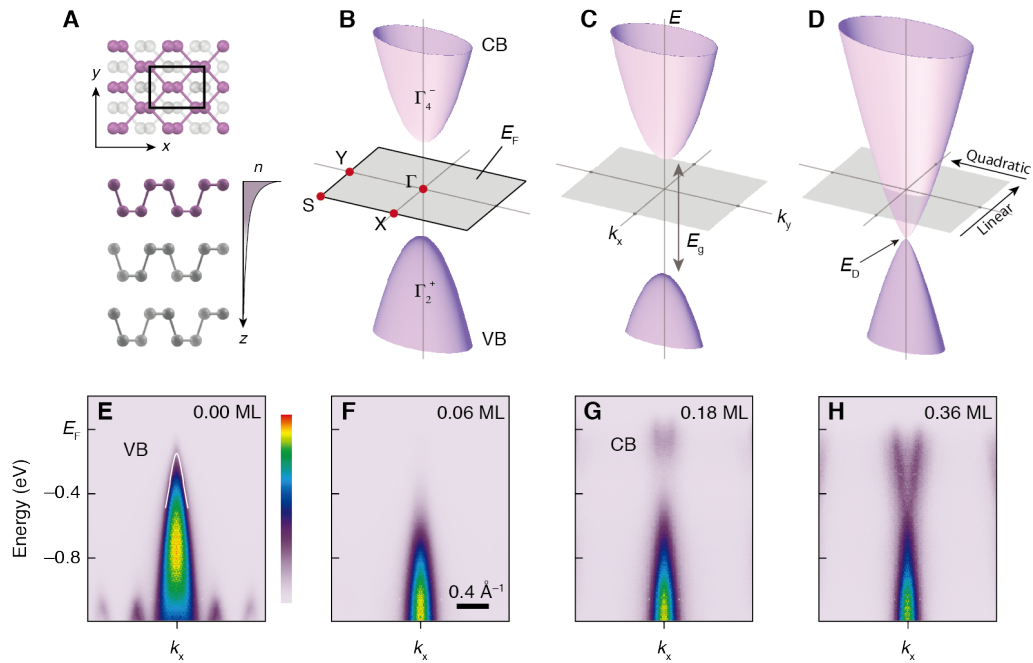

**Fig. 1. Band structure and tunable $E_g$ of few-layer black phosphorus.** (**A**) Atomic structure of BP. The solid square is the surface unit cell, and the interlayer distance is about 5 Å. Lower right: density profile of 2D electron gas decaying along $z$ towards the bulk. Band structure of (**B**) pristine BP, (**C**) BP in the initial stage of surface doping, and (**D**) BP at the transition to a zero-gap semimetal. The solid square in (B) is the surface Brillouin zone with high symmetry points marked by red circles. (**E** to **H**) Experimental band structure of BP taken at 15 K near $E_F$ along $k_x$ with dopant density marked at the upper right. The photon energy is 104 eV for $k_z$ at the Z point of the bulk Brillouin zone (*29*). The dopant density is estimated in unit of monolayer (ML) from simultaneously taken K $3p$ core-level spectra (fig. S7). The white line overlaid in (E) is a fit to VB with the $k$-$p$ perturbation formula (*31*).

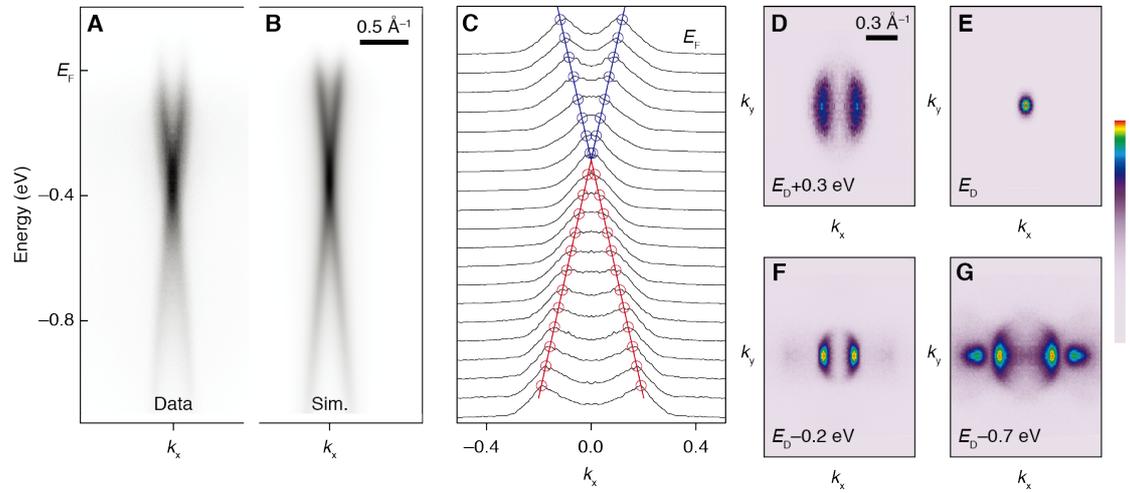

**Fig. 2. Anisotropic Dirac semimetal state at the critical density.** (**A**) High-resolution ARPES data taken at 15 K along the armchair direction $k_x$. (**B**) Corresponding spectral simulation with two linearly crossing bands and finite broadening (*31*). (**C**) Normalized momentum-distribution curves (0.05 eV steps from $E_F$) with their peak positions marked by open circles. Red and blue lines are linear fits to VB and CB, respectively. (**D** to **G**) Series of ARPES intensity maps at constant energies (marked at the bottom), shown over a whole surface Brillouin zone (Fig. 1B).

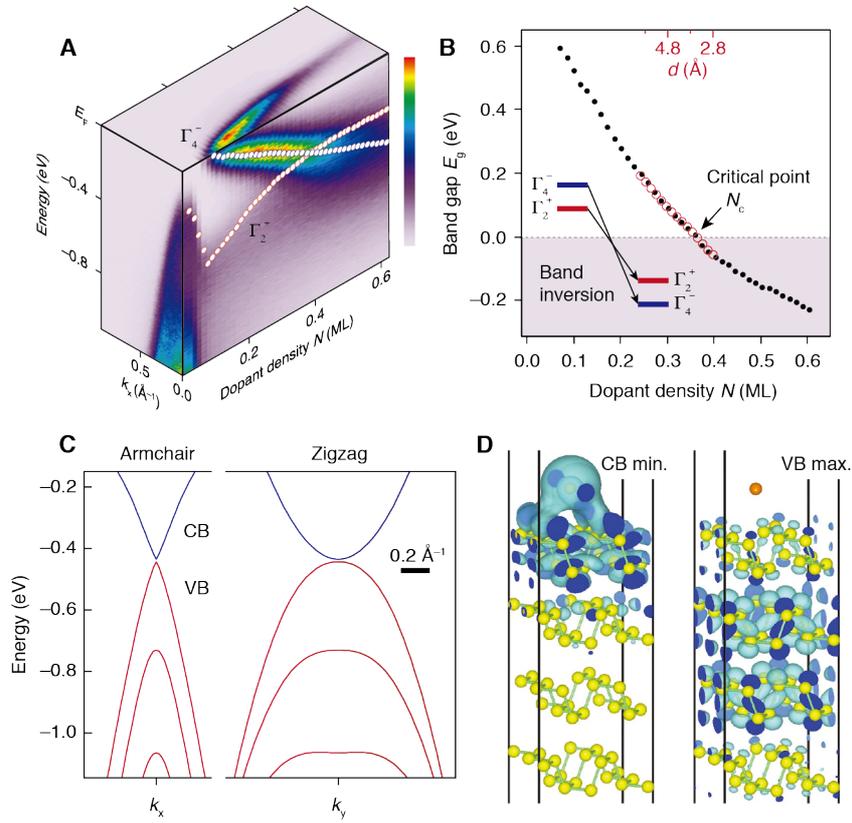

**Fig. 3. Quantitative analysis for the control mechanism of $E_g$.** (**A**) 3D representation of the band evolution as a function of $N$, taken at 15 K. Red diamonds and blue circles are positions of $E_v$ and $E_c$, respectively. (**B**) $E_g$ as a function of $N$. The negative gap in the shaded region represents the inverted gap at the G point. Overlaid red circles are $E_g$ from DFT band calculations where varying dopant density is simulated by changing the vertical distance $d$ between K and BP (*31*). (**C**) Theoretical band dispersions for K-doped BP films at $d = 3.67$ Å. (**D**) Partial charge densities of $E_v$ and $E_c$ points in (C), separated by a tiny gap less than 10 meV. The light blue shows the isosurface, set to $1.76 \times 10^{-3}/\text{Å}^3$, and its cross-sections are shown by the dark blue. Yellow and orange balls represent P and K atoms, respectively.

## Materials and Methods

### Experiments and samples

ARPES measurements were conducted at Beamline 4.0.3 in Advance Light Source and Beamline 4A1 in Pohang Light Source II, equipped with R8000 and R4000 analyzers (VG-Scienta, Sweden), respectively. Data were collected at 15−150 K. Energy and momentum resolutions were better than 20 meV and 0.02 Å$^{-1}$. The photon energy 104 eV was chosen for $k_z$ at the Z point of the bulk Brillouin zone (*29*). The photon-energy dependence was checked in the range of 85 ~ 104 eV, which is over half the bulk Brillouin zone along $k_z$ (gray arrow, fig. S1). Single-crystal BP samples (Smart-Elements) were cleaved at low temperature in the ultrahigh vacuum chamber with the base pressure of 3−5 × 10$^{-11}$ torr. The deposition of K atoms was done in-situ by a commercial (SAES) getter source. The K density was calibrated in unit of ML from K 3*p* core-level spectra (fig. S7), taken together with valence-band spectra.

### *k-p* theory and effective mass

The *k-p* theory provides an approximate analytic scheme, widely used to model the band dispersion of narrow-gap semiconductors (*35*). At around a high-symmetry point, where $E_v$ and $E_c$ are separated by direct $E_g$, the perturbation term describes *k*-dependent interband mixing, and gives a linear component in band dispersion as

$$E(k) = E_0 \mp E_g/2 + \hbar^2 k^2/2m_e \pm \sqrt{(E_g/2)^2 + (P \cdot k)^2}$$

where $E_0 = (E_v + E_c)/2$, $m_e$ is the electron rest mass, $\hbar$ is the Planck constant divided by 2π, and $P$ is the optical matrix element (*35*). As $E_g$ approaches zero, the dispersion becomes linear with $E(k) \sim P \cdot k$, and the effective mass is given by

$$m^* = \hbar^2 E_g / (2P^2)$$

The parameters for our calculations, $E_g$ = 0.335 eV and $E_0$ = 0.018 eV, were taken from the previous reports (*36*). The best fit to VB (white line in Fig. 1E) yields the hole effective mass along $k_x$ (the light-mass direction) as $m_x^*$ = 0.08 ± 0.03$m_e$, which is in good agreement with theoretical calculations (*14, 36*).

### Spectral simulations

Our model is based on the standard spectral-function formula in a Lorenzian form as

$$I(E,k) \propto \frac{\sigma(E)}{(E - E_b(k))^2 + \sigma(E)^2} \cdot f_{FD}(E)$$

where $E_b(k)$ is the band dispersion, σ(*E*) is the spectral width, $f_{FD}(E)$ is the Fermi-Dirac function. $E_b(k)$ is taken from linearly crossing bands as ±$v_x k$, where the band velocity is $v_x$ = 5.1 × 10$^5$ m/s. σ(*E*) is set by energy resolution, 30 meV at $E_F$ from which it monotonically increases with binding energy. This increment with binding energy is to take into account the self-energy effect. $E_F$ is determined by experimental data in Fig. 2A.

Band calculations

We first determined the equilibrium lattice constants and internal parameters of bulk BP (Space group number: 64, Abma) using the WIEN2k code (*37*), obtaining the equilibrium lattice constants of $a = 4.57$, $b = 3.32$, and $c = 11.33$ Å. We used the generalized gradient approximation (GGA) (*38*) to the exchange-correlation energy, and obtained the $E_g$ of 0.21 eV for bulk BP and 0.98 eV for phosphorene. These values agree well with the previous experimental and theoretical reports (*8, 9*). The optimization of cell parameters and atomic positions is important in achieving a finite $E_g$ without hybrid functionals or GW approximations (*13*), since the $E_g$ of BP strongly depends on the lattice constant along the *c*-axis (*17, 18, 22*). As shown in fig. S8, we confirmed that the overall band structure with GGA (PBE) is in good agreement with experimental ARPES data (fig. S2) as well as that with hybrid functionals (*8*). Then, using the optimized bulk structure, we constructed multilayer BP structures, and calculated their band structures using the SIESTA code (*39*), where the exchange-correlation was treated with GGA. In our calculations, electronic wavefunctions and charge densities were projected onto a real-space grid with an equivalent energy cutoff of 500 Ry, and they are integrated using a 10×10 *k*-grid in the 2D Brillouin zone for all multilayer-BP calculations.

For K-doped BP, a single K atom was added on each 2×2 surface unit cell of multilayer BP (corresponding to $N \sim 0.4$ ML in experiments). Atomic relaxations were performed for the K atom and the topmost BP layer, and atomic displacements of P atoms were found less than 0.1 Å, showing that the strain effect on the variation of $E_g$ (*15−17*) is negligible in our K-doped BP systems. In order to effectively describe a lower K density, the vertical distance *d* between K and BP at equilibrium (2.76 Å) is gradually increased to reduce the effect of K atoms without change in the supercell size. The validity of this method was checked by comparison to results from larger supercell calculations with lower K densities at equilibrium *d*. As for the field dependence of $E_g$, we note that the previous PBE and HSE06 calculations have proven to show a fully consistent behavior (*17, 19, 21*). That is, the overall trend of reduction in $E_g$ as a function of electric field is regardless of the absolute value of $E_g$ at the starting point (that of pristine BP).

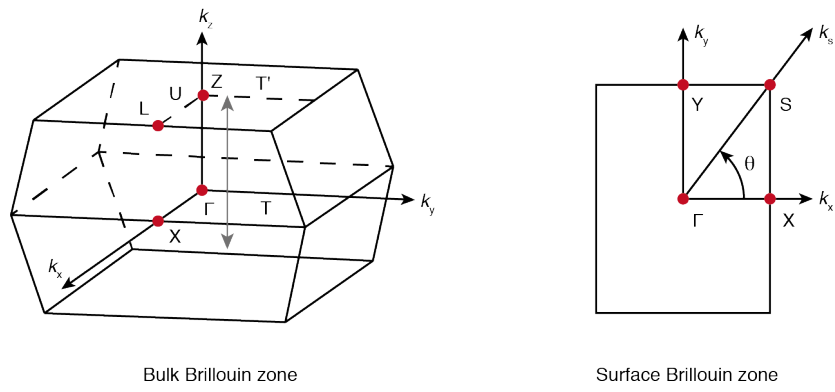

**Fig. S1. Brillouin zone of BP.** 3D bulk Brillouin zone (left) and corresponding 2D surface Brillouin zone (right). The high-symmetry points are marked by red circles.

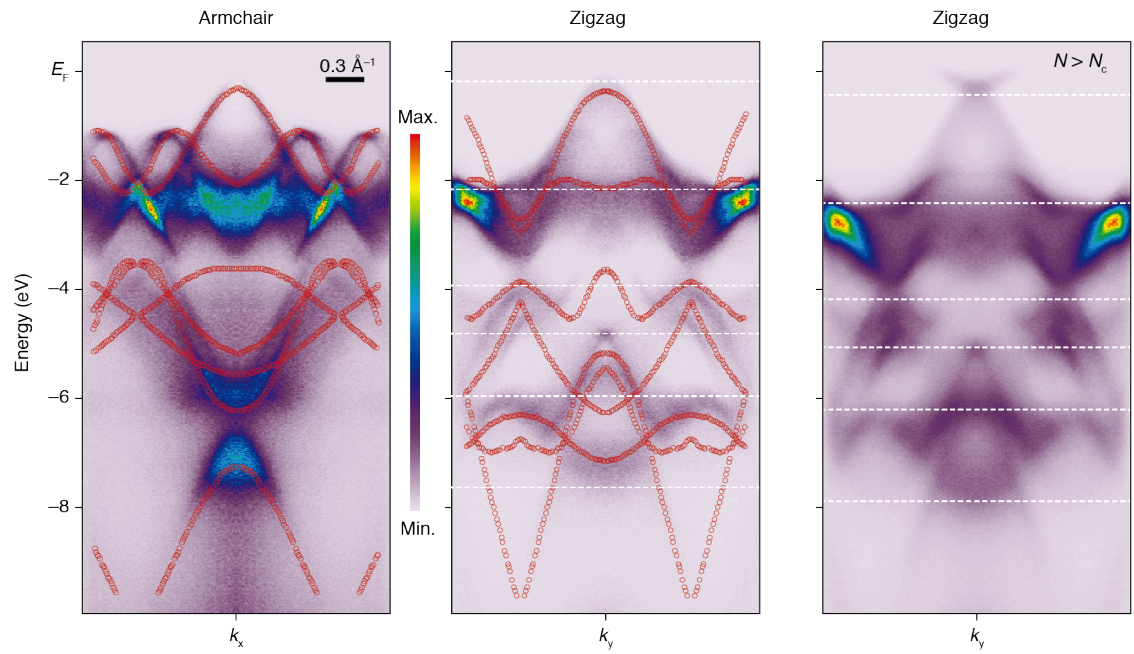

**Fig. S2. Wide-scale band structure and rigid band shift with doping.**
Experimental band structure of pristine BP, measured by ARPES at 15 K along armchair ($k_x$, left) and zigzag ($k_y$, middle) directions with the photon energy of 104 eV. Overlaid red circles are those from band calculations using the self-consistent pseudopotential method (*36*), which is overall in good agreement. Shown in the right panel is that measured along the zigzag direction ($k_y$) after doping at 0.42 ML ($N > N_c$), confirming the rigid band shift over the wide energy.

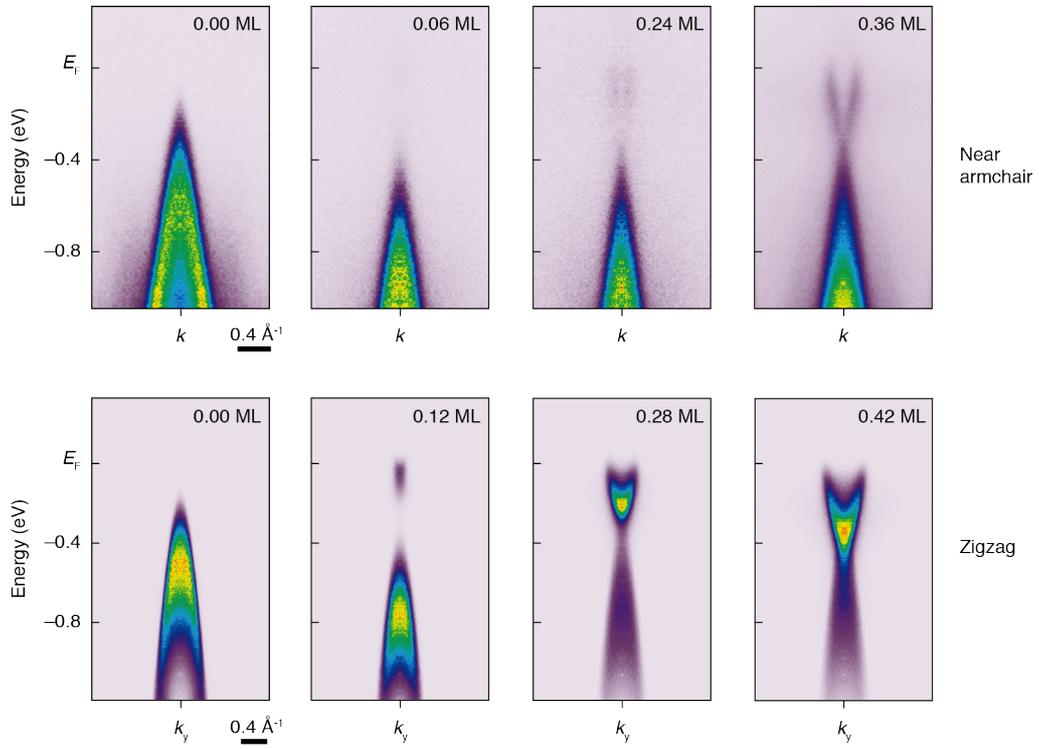

**Fig. S3. Doping evolution of bands along high-symmetry directions.** Series of ARPES spectra with $N$, taken from two different samples along near-armchair and zigzag directions (*30*), as shown in fig. S1. Each dopant density is marked at the upper right.

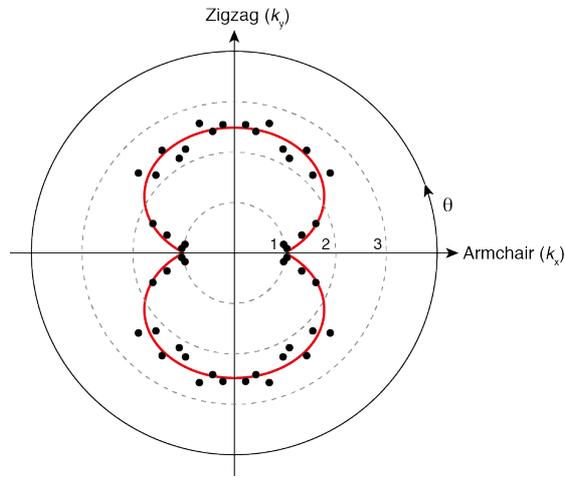

**Fig. S4. Polar plot of band dispersion at the critical density.** The dispersion of CB is fit by a simple power-law function, and the power value ($p$) is plotted as a function of in-plane angle ($\theta$), as shown in fig. S1. This diagram shows how linear dispersion ($p = 1$) along the armchair direction ($k_x$) changes to quadratic dispersion ($p \sim 2$) along the zigzag direction ($k_y$). The red line is a guide for ellipsoidal anisotropy.

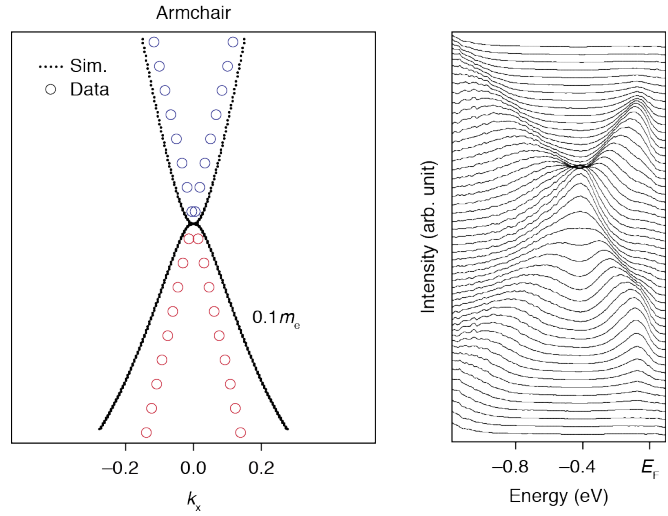

**Fig. S5. Effective mass of crossing bands at $N_c$.** The ARPES peak positions shown in Fig. 2C are directly compared to the two quadratic bands of $0.1m_e$, simulated by the *k-p* formula and shifted to have zero $E_g$. The error bar of peak positions is smaller than the size of open circles. This clearly confirms that, although the $m_x^*$ of pristine BP is very small, the rigid shift of VB and CB cannot account for the experimental data. Shown on the right are corresponding energy distribution curves of crossing bands, which supports the absence of $E_g$ at the crossing point.

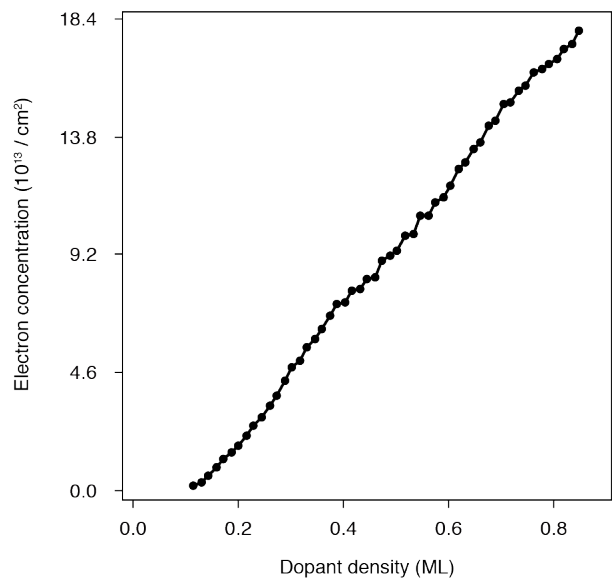

**Fig. S6. Electron concentration as a function of dopant density.** Electron concentration is estimated from the area enclosed by the Fermi surface, after CB drops below $E_F$ by doping. This linear relation indicates a monotonic charge transfer from dopants to surface phosphorene layers.

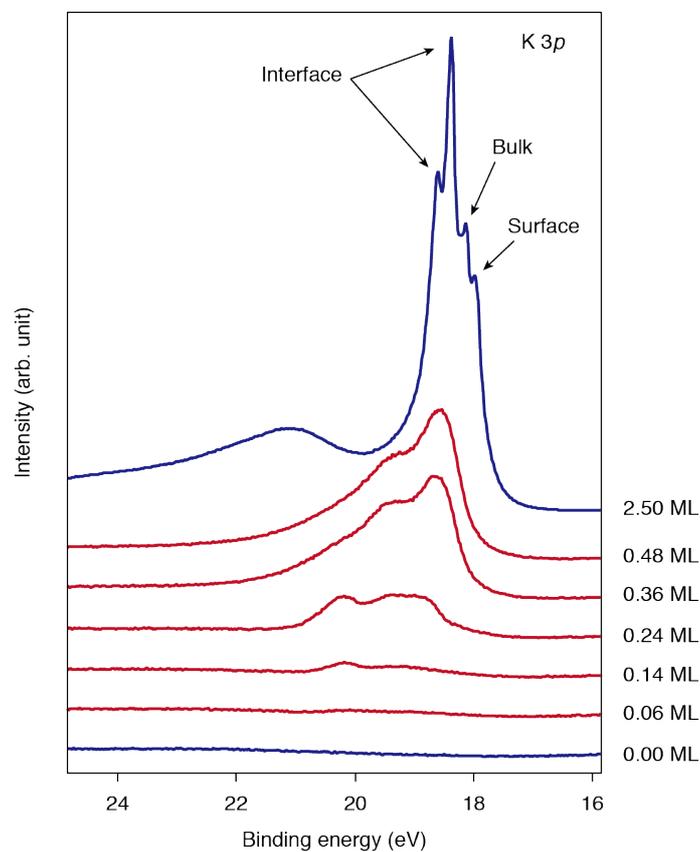

**Fig. S7. Potassium coverage calibration.** Evolution of K 3*p* core-level spectra with the deposition of K atoms. The deposition rate was calibrated based on characteristic features at 1 ~ 3 ML, previously reported in the epitaxial growth on several different substrates (*40, 41*). The coverage lower than 1 ML could be estimated from a relative areal ratio of core-level spectra after subtracting the background (that of pristine BP at 0.00 ML).

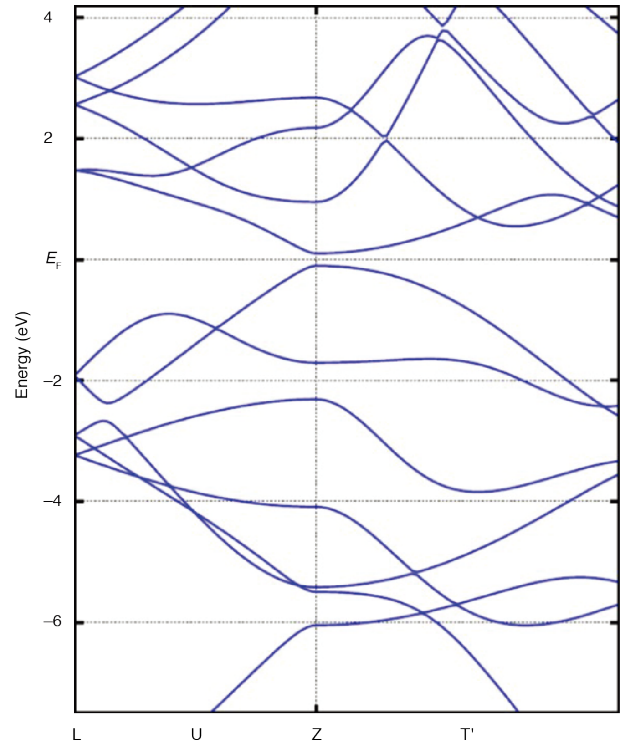

**Fig. S8. Calculated band structure of bulk BP.** The GGA (PBE) band calculations with fully relaxed cell parameters and atomic positions (given in the materials and methods section) are in good agreement with our experimental ARPES data (fig. S2) as well as HSE06 calculations (*8*). The high symmetry points of the bulk Brillouin zone are shown in fig. S1.

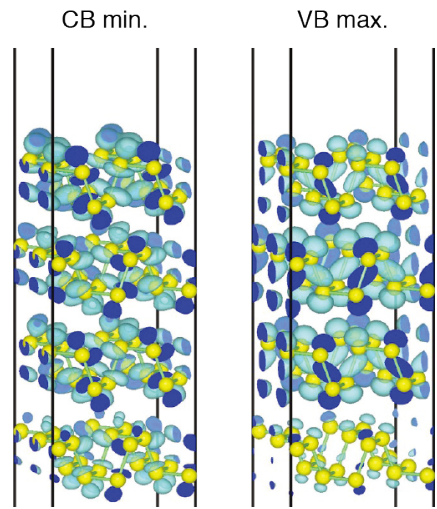

**Fig. S9. Partial charge density plot of four-layer BP.** The $E_v$ and $E_c$ states are uniformly distributed over the BP layers. The light blue shows the isosurface that is set to $1.76 \times 10^{-3}/\text{Å}^3$, and its cross-sections are shown by the dark blue. Yellow balls represent P atoms.